\documentclass[twocolumn,aps,prb]{revtex4}
\usepackage{graphics,graphicx}
\usepackage{dcolumn}
\usepackage{bm}

\newcommand{\Ref}[1]{Ref.~\onlinecite{#1}}

\def\eb{\begin{equation}}   
\def\ee{\end{equation}}     
\def\ea#1{\begin{eqnarray} #1 \end{eqnarray}}   

\def\shro{Schr\"odinger}

\def\ra{\rightarrow}

\def\im{\text{Im}}
\def\re{\text{Re}}

\def\of#1{\left(#1\right)}

\def\eq#1{Eq.~(\ref{#1})}
\def\eqs#1#2{Eqs.~(\ref{#1}) and (\ref{#2})}

\def\sof#1{\left[ {#1} \right]}
\def\bof#1{\left\{ {#1} \right\}}

\def\Dlt{\Delta}

\def\Pp{\Psi_+}
\def\Pm{\Psi_-}
\def\Ppm{\Psi_\pm}
\def\Pmp{\Psi_\mp}

\def\tshift{t_{\text{shift}}}
\def\tmax{t_{\text{max}}}
\def\Veff{V_{\text{eff}}}

\begin{document}

\title{Development and Numerical Analysis of \\
``Black-box'' Counterpropagating Wave Algorithm for \\
Exact Quantum Scattering Calculations}

\author{Bill Poirier}
\affiliation{Department of Chemistry and Biochemistry, and
         Department of Physics, \\
          Texas Tech University, Box 41061,
         Lubbock, Texas 79409-1061}
\email{Bill.Poirier@ttu.edu}

\begin{abstract}

In a recent series of papers [J. Chem. Phys. {\bf 121} 4501 (2004),
J. Chem. Phys. {\bf 124} 034115 (2006), J. Chem. Phys. {\bf 124}
034116 (2006)] a bipolar counter-propagating wave decomposition,
$\Psi = \Psi_+ + \Psi_-$, was presented for stationary bound states
$\Psi$ of the one-dimensional \shro\ equation, such that the components
$\Ppm$ approach their semiclassical WKB analogs in
the large action limit. The corresponding bipolar quantum trajectories
are classical-like and well-behaved, even when $\Psi$ has many nodes, or
is wildly oscillatory. In this paper, the earlier results are used to
construct a universal ``black-box'' algorithm, numerically robust,
stable and efficient, for computing accurate scattering quantities of
any quantum dynamical system in one degree of freedom.

\end{abstract}

\maketitle



\section{INTRODUCTION}
\label{intro}

In a recent series of three
articles,\cite{poirier04bohmI,poirier06bohmII,poirier06bohmIII}
the use of bipolar counter-propagating wave methods
(CPWMs)\cite{poirier04bohmI,poirier06bohmII,poirier06bohmIII,babyuk04,wyatt}
for exact numerical solution of the time-independent \shro\ equation was
explored. The basic underlying idea is to decompose the stationary wavefunction, $\Psi$,
into a sum of two counter-propagating wave components, i.e.
\eb
     \Psi = \Pp + \Pm.  \label{psitot}
\ee
If done appropriately, such a decomposition can lead to very important
ramifications for quantum trajectory methods
(QTMs)\cite{wyatt,lopreore99,mayor99,wyatt99,shalashilin00,wyatt01b,wyatt01c,burghardt01b,bittner02b,hughes03}---i.e.,
trajectory-based numerical techniques for performing exact quantum dynamics
calculations, based on Bohmian
mechanics\cite{madelung26,vanvleck28,bohm52a,bohm52b,takabayasi54,holland}---due
to nonlinearity of the Bohmian equations of motion. In particular, the earlier
series of articles has culminated in a set of trajectory-based time-dependent
methods for computing stationary scattering quantities
(the theoretical underpinning of all chemical reactions)
in one degree of freedom (DOF). These methods were found to be numerically
stable and efficient for model systems within a certain regime of system
parameters and accuracy typical of molecular
applications.\cite{poirier06bohmII,poirier06bohmIII} However,
more recent investigations have uncovered certain limitations that arise in
more extreme cases. This has motivated the goal of the present paper, to perform a much
more detailed numerical analysis of the methods, and to develop a single stable,
robust, and efficient ``black-box'' algorithm that can be applied to virtually
any 1 DOF system with a minimum of user intervention.

In recent years, QTM's have arisen as a very promising tool for performing
accurate quantum dynamics calculations for many-DOF systems, using trajectory
ensembles in a manner similar to classical simulations.\cite{frenkel}
In the conventional unipolar formulation of Bohmian
mechanics,\cite{madelung26,vanvleck28,bohm52a,bohm52b,takabayasi54,holland}
the Madelung-Bohm amplitude-phase decomposition of the 1 DOF $\Psi$, i.e.
\eb
\Psi(x,t) = R(x,t) e^{i S(x,t) /\hbar}, \label{ansatz}
\ee
is used to generate the (presumably classical-like) quantum field functions,
$R(x,t)$ and $S(x,t)$. For slowly-varying potentials, the
classical field functions are also slowly-varying, except in the vicinity of
caustics. The corresponding quantum field functions of \eq{ansatz} behave
similarly when there is no interference. However, interference
introduces non-classical-like oscillations in the quantum field functions,
which in turn lead to numerical difficulties for QTM calculations---collectively
referred to as ``the node problem''\cite{wyatt,wyatt01b} (even though true nodes
{\em per se} need not be present).  As interference is formally unavoidable in any
reactive scattering context, the node problem is to date the most important
issue impeding the progress of QTM's as a general and robust tool for scattering
dynamics applications, although much progress has been
made.\cite{babyuk04,trahan03,kendrick03,pauler04}

The bipolar CPWM approach to the node problem, as adopted here,
and in the earlier series of
articles,\cite{poirier04bohmI,poirier06bohmII,poirier06bohmIII}
is inspired by classical and semiclassical theories---which employ
a multipolar, \eq{psitot}-like decomposition as the means of preserving
smooth and slowly-varying field functions. For stationary 1 DOF applications,
the $\Psi_+$ and $\Psi_-$ components are respectively, left- and
right-traveling waves, with equal and opposite (group) velocities, $v(x)$
and $-v(x)$. Oscillatory interference behavior is not evident in the individual
component field functions, but arises naturally from their linear superposition.
For stationary bound state calculations, cancellation of flux requires that the
component densities, $\rho_+(x) = |\Psi_+(x)|^2$ and $\rho_-(x) = |\Psi_-(x)|^2$,
be equal. A reasonable generalization of classical field properties can
be used to arrive at a unique exact quantum decomposition of the \eq{psitot}
form, which---unlike the conventional unipolar ansatz [\eq{ansatz}]---satisfies the
correspondence principle in the large-action limit.\cite{poirier04bohmI}

For stationary scattering, or continuum state calculations, the non-uniqueness
and non-square-integrability of the quantum solutions complicate matters
somewhat,\cite{poirier06bohmII,poirier06bohmIII} as might be expected.
Boundary conditions play an essential role,
and specifically for the usual left-incident boundary conditions presumed here,
necessarily imply asymmetry and non-zero flux for the corresponding exact quantum
stationary solution.  Conventional scattering theory already provides an {\em asymptotic}
(i.e. $|x| \ra \infty$) bipolar decomposition into the familiar ``incident,''
``transmitted,'' and ``reflected'' plane wave components---in fact the only
decomposition possible that does not result in asymptotically oscillatory
field functions. An exact quantum bipolar CPWM decomposition
that respects all of the above can be
constructed,\cite{poirier06bohmII,poirier06bohmIII}  with $\Psi_+(x)$ joining
the incident and transmitted waves continuously through the interaction region,
and $\Psi_-(x)$, the reflected wave, damping to zero as $x \ra \infty$. This is
again achieved through analogy with semiclassical mechanics, albeit a
``sophisticated'' version.\cite{heading,froman,berry72} A key feature
is that the trajectories are actually {\em classical}, with all
quantum effects arising not from a quantum potential (as in conventional
Bohmian mechanics), but through dynamical $\Ppm$ coupling. Conceptually,
this gives rise to instantaneous trajectory {\em hopping}\cite{tully71}
from one CPWM component to the other, in the interaction region where the
potential $V(x)$ is changing. In effect, one has a {\em local, time-dependent}
theory of stationary scattering, instead of the more traditional
{\em global, time-independent} picture.

The new theory thus provides some pedagogical insight, analogous to ray optics
and cavity ring-down experiments.\cite{poirier06bohmII,wheeler98}
However, this paper focuses primarily on numerical aspects of the
new approach---which may be regarded as a relaxation
method with exponentially-fast convergence, for computing stationary
scattering states of desired boundary conditions, and associated
reactive scattering quantities such as reflection
and transmission probabilities.
In comparison with other quantum scattering methods, the bipolar CPWM approach
offers some decided numerical advantages. In particular, there is no
need to invoke complex scaling
(analytical continuation)\cite{complexscaling78,reinhardt82,ryaboy94}
or absorbing boundary conditions
(optical potentials),\cite{jolicard85,seideman92a,seideman92b,riss93,poirier03capII,poirier03capI,muga04}
as trajectories and field quantities are real-valued throughout.
Moreover, the scaling of computational (CPU) effort is linear with
the grid size, $N$, rather than proportional to $N^3$.

The new approach may therefore lead to some of the most efficient exact
quantum scattering algorithms in existence, but certain practical
limitations must first be overcome. To begin with, the previous papers
propose several distinct algorithms; a detailed numerical analysis must
be performed to determine which of these is ``best,'' from the perspective
of numerical stability, robustness, and efficiency. In particular, some
of these algorithms appear to be ineffective beyond a certain desired
level of accuracy (Sec.~\ref{oldanalysis}). The reasons for this must
be elucidated via systematic study, in order to develop more efficient
implementations. An even greater motivation, though, is the fact that the
above problem is greatly exaggerated in the multidimensional case, resulting in
numerical instabilities that have thus far hindered efforts in this direction.
Another limitation is that some of the previous algorithms
are not applicable to completely general systems
(Sec.~\ref{oldmethods}). This is addressed here by generalizing the
methodology to allow for a nearly arbitrary choice of ``classical''
trajectory (Sec.~\ref{general}).  Finally, the robustness of the
new algorithms must be demonstrated by application to representative
molecular systems encompassing a very broad range of potentials,
system parameters, and desired accuracy levels. One such application area,
known to cause difficulties for conventional semiclassical and exact quantum
methods, is particularly important---i.e., the deep tunneling regime
(Sec.~\ref{deeptunnel}). Other examples considered here include asymmetric
potentials with barriers (Sec.~\ref{barrier}) and systems with
reaction intermediates (Sec.~\ref{double}).


\section{BACKGROUND}
\label{background}

\subsection{Exact Quantum Dynamics Using Counterpropagating Trajectories}
\label{oldmethods}

Here, we summarize some of the developments of Refs.~1-3. The basic
goal is to define a set of time-evolution equations for the two 1 DOF
counterpropagating waves, $\Ppm(x,t)$, such that in the large $t$ limit,
the total $\Psi(x,t)$ of \eq{psitot} approaches the exact stationary state
evolution for the desired energy, $E$, and boundary conditions.  This general
approach is thus a ``relaxation method,'' for which $\Psi(x,t)$ at intermediate
times need not adhere to the actual time-dependent \shro\ equation---although
there are versions of the method for which the latter property is also
satisfied in a sense.\cite{poirier06bohmII,poirier06bohmIII}
For the usual left-incident boundary condition, the initial reflected
wave must be zero, i.e. $\Pm(x,0) = 0$.
The initial $\Pp(x,0)$ is essentially arbitrary, as it is
presumed\cite{poirier06bohmII,poirier06bohmIII}
that any reasonable choice will converge
exponentially quickly in the large time limit. However, two natural choices
have been considered: (1) $\Pp(x,0) = \exp\!\sof{ i \sqrt{2 m E} x/\hbar}$,
i.e. incident asymptotic plane wave extended throughout space
[with $V(x) \ra 0$ presumed as $x\ra -\infty$];
(2) $\Pp(x,0) = \Theta(x_L-x) \exp\!\sof{ i \sqrt{2 m E} x/\hbar}$,
i.e. incident wave truncated at left edge of interaction region,
$x_L \leq x \leq x_R$.  In \Ref{poirier06bohmIII}, (1) and (2)
above are referred to respectively as  ``non-wave-front'' and ``wave-front''
versions. The latter gives rise to a useful cavity ring-down dynamical
interpretation,\cite{poirier06bohmII,wheeler98} although the former is found
to converge more quickly to the exact solution, and is therefore used
throughout this paper. Note that (1) also has its own interpretation,
to be discussed shortly.

From \Ref{poirier06bohmIII}, we also learn that there are two distinct semiclassical
formalisms that can be applied to generate exact \eq{psitot} decompositions
(and associated time-evolution equations) with the desired theoretical
properties---the Bremmer approach (B)\cite{berry72}
and the Fr\"oman approach (F).\cite{froman} Of
the two, the B approach can only be applied when the
trajectories are classical trajectories.
Unfortunately, classical trajectories cause numerical instabilities for both
B and F evolutions whenever there is tunneling, i.e. for all below-barrier
energies, $E$, and is therefore not a viable choice for a completely ``robust''
algorithm. The F approach, on the other hand, can be generalized for
almost {\em any} desired choice of ``semiclassical'' trajectory. In particular,
constant velocity trajectories (where the speed corresponds to the incident
plane wave energy $E$) handle tunneling without difficulty, and lead to
smoother field functions and remarkably simple F time-evolution equations:
\eb
     {d \Ppm \over dt} = {i \over \hbar} \of{E-V}\Ppm -
                         {i \over \hbar} V \Pmp \label{Pdotunigwf}
\ee

Some comments on \eq{Pdotunigwf} are in order. First, these are the
Lagrangian, or ``hydrodynamic'' equations of motion, meaning that the
left hand-side is a total time derivative, as would be appropriate
for a trajectory (moving-grid) calculation. Note that they require
no differentiation of any kind---not even of the potential itself,
to construct force fields. Note also that the final, coupling term,
i.e. that responsible for all quantum corrections beyond the
``semiclassical'' approximation, is proportional to $V(x)$, and
vanishes in the $x\ra -\infty$ reactant asymptotic limit.
It also vanishes in the product asymptote provided the potential is
``asymptotically symmetric,'' i.e.
$V(x) \ra = 0$ as $x\ra \infty$ (the asymmetric case will be
addressed in Sec.~\ref{discontinuous}). This ensures that the $\Ppm(x)$
solutions exhibit the requisite plane wave behavior for the desired
energy $E$. Finally, we note that for the special case $V=0$,
i.e. the potential that would classically give rise to the constant
velocity trajectories used, all coupling vanishes. This implies
that the non-wave-front initial condition (1) described above is in
fact the ``basic WKB''\cite{poirier06bohmIII,froman} approximation
in the generalized F sense, as will be exploited in Sec.~\ref{general}.

Since the $\Pp$ and $\Pm$ trajectories move in opposite directions,
the two moving grids are necessarily incommensurate at an arbitrary time,
which numerically necessitates the use of interpolation for evaluating
the coupling term in \eq{Pdotunigwf}. Alternatively, an ``Eulerian''
or fixed-grid implementation can be adopted, for which the corresponding
time-evolution equations are
\eb
     {\partial \Ppm \over \partial t} = \mp v \Ppm' +
                         {i \over \hbar} \of{E-V}\Ppm -
                         {i \over \hbar} V \Pmp \label{Pdotuni},
\ee
where $v = \sqrt{2 E/m}$ is the trajectory speed, and primes denote spatial
differentiation. The grids are now commensurate at all times, obviating
the need for interpolation, but numerical differentiation is now
required to evaluate the new convective term in \eq{Pdotuni}.

\subsection{Numerical Analysis of Fixed-grid Scheme}
\label{oldanalysis}

Of the various possible numerical schemes presented in \Ref{poirier06bohmIII} and
above, we have thus far settled on just two candidates for the present purpose:
the trajectory and fixed-grid versions of the non-wave-front, F,
constant-velocity trajectory method. To choose one over the other, we must
consider the stated criteria of numerical stability, efficiency, and robustness
(i.e. generality of applicability across different systems and computational
accuracy levels).

In \Ref{poirier06bohmIII}, all of the calculations were applied to one degree-of-freedom
(1 DOF) systems with similar parameter values and energies, and for the most part,
to a desired accuracy level of $10^{-4}$.  In this context, the
two numerical schemes were found to be of roughly comparable efficiency
(i.e. total CPU time), with the trajectory approach requiring substantially fewer
grid points (around $4\times$ per DOF) and larger time steps ($\times 10$),
but a much longer CPU time per step, due primarily to the nearest-neighbor
search algorithm that was employed. On the other hand, there were also
clear indications that the trajectory method would become superior at higher
accuracies---e.g., \Ref{poirier06bohmIII} Fig.~5, which suggests that the fixed-grid
error for reflection and transmission probabilities cannot be reduced below
$10^{-5}$ or so, no matter how many grid points are used.

A limit on achievable accuracy is clearly not a desirable trait for a robust
method, and in particular, is complete anathema for accurate deep tunneling
applications. Consequently, a detailed numerical analysis of this situation was
performed for this paper, in order to gain a thorough understanding of the
factors responsible. In the course of this, it was discovered that
the limiting accuracy can be substantially worse for other reasonable system
parameter choices, especially for multidimensional systems, and/or those with reaction
intermediates---lending additional imperative to the exercise. The manifestation
of the error was quickly identified as oscillations in the right-asymptotic
density $\rho_+(x) = |\Pp(x)|^2$, which form around $x=x_R$, and then propagate
inward over time---whereas it is clear from \eq{Pdotunigwf} that the solution
$\rho'_+(x)$ should be zero asymptotically.

\Ref{poirier06bohmIII}, and the above-mentioned preliminary numerical investigations
for this paper, computed fixed-grid spatial derivatives using centered,
fourth-order finite
differences\cite{fornberg88} (one-sided for the endpoints $x_R$ and $x_L$),
and fourth-order Runge-Kutta time integration.\cite{phillips}
However, the bulk of the present analysis uses second-order finite difference
(two-sided for interior grid points and one-sided for endpoints) and first-order
forward Euler time integration, as a deliberate means of exacerbating the problem
and simplifying the analysis, in order to identify the cause.
The oscillation amplitudes were indeed found to be
much larger in this case. It was also found that the oscillation
wavelength was always 2-3 grid points, regardless of the density of grid
points used. With respect to number of grid points, $N$, (keeping $x_L$ and
$x_R$ fixed), the amplitude oscillation was found to decrease roughly inversely.
Reducing the time step had no effect whatsoever on the oscillation amplitude
or wavelength, but did reduce the inward propagation speed, roughly
proportionally.

All of the above findings support the notion that the oscillation problem
is somehow due to spatial differentiation. It might also be theorized that
oscillations arise from the $\Ppm$ phase difference inherent in the coupling term
of \eq{Pdotunigwf}, but the wavelengths are incommensurate with what is actually
observed. To definitively rule out coupling as the source
of the problem, however, a calculation was performed for the $V=0$ free
particle system, for which there is no coupling---yet the above oscillations
were still observed. Another alternative explanation would be von-Neumann
instabilities.\cite{press} A straightforward theoretical analysis is
possible for the free particle case, and so this was conducted, and compared
to numerical tests for a wide range of time steps and grid spacings. In
the theoretically von Neumann stable regime, the oscillations were still observed,
whereas in the von Neumann unstable regime, the numerical solutions diverged
over time as expected---a different phenomenon entirely.

In fact, {\em none} of the vast literature on PDE simulations
of which the author is aware can account for the odd oscillatory behavior observed.
The only sensible explanation of the cause can therefore be the ``mixed''
boundary conditions, which are highly non-standard, and perhaps have not been
considered previously. In particular,
\eb
     \Pm(x_R,t) = 0 \qquad ;
     \qquad \Pp(x_L,t) = \exp\sof{- {i \over\hbar} E t},
     \label{bcs}
\ee
so that the two different components have boundary conditions on
opposite sides. This situation is completely natural, given the
counterpropagating nature of the method, and cannot be changed through any
straightforward theoretical recasting. Within the fixed-grid framework,
various numerical solutions have been essayed, including asymptotic plane
wave rescaling, and amplitude/phase decomposition of the endpoint $\Ppm$
components, but none appears to provide substantial relief from the oscillation
problem described above.


\section{THEORY AND ALGORITHM DEVELOPMENT}
\label{theory}

\subsection{Constant-velocity Trajectory Scheme}
\label{constant}

We are thus led naturally to the trajectory, rather than fixed-grid,
version of the non-wave-front, F, constant-velocity method. \Ref{poirier06bohmIII}
Fig.~5 indicates the high accuracies (around $10^{-8}$) of which the
trajectory approach is capable. This can be largely attributed to the
fact that the trajectory version does not require numerical spatial
differentiation. However, a more detailed numerical analysis reveals
that the errors in this case also manifest as asymptotic oscillations,
albeit of much smaller amplitude than the fixed-grid case, and with
different scaling. More specifically, for a test Eckart system
comparable to that of \Ref{poirier06bohmIII}, an inverse quartic relation is
found between the oscillation amplitude and grid size, $N$. This can also
be roughly inferred from \Ref{poirier06bohmIII} Fig.~5, and is in any case
not surprising, given that third-order interpolation is
used (Sec.~\ref{interpolation}).

As in the fixed-grid case, the oscillation amplitudes and wavelengths
are completely unaffected by time-step size. However, the latter quantity
does turn out to be closely related to another, more serious type of
numerical error also observed: in the large $t$ limit, the magnitude
of $\Pp(x,t)$ grows linearly with $t$ in the left-asymptotic region.
For typical $\Dlt$ values
and desired accuracy levels, using fourth-order time evolution as in
\Ref{poirier06bohmIII}, the growth rate is so slow as to contribute insignificantly
to the final error. However, this asymptotic growth does present a very
significant challenge for extremely accurate calculations (e.g. $10^{-12}$
as in Sec.~\ref{deeptunnel}) and must therefore be addressed. Once again,
to exaggerate the effect for analytical purposes, first-order forward Euler
time integration was used. For the test Eckart system, the growth rate (with
respect to system time, {\em not} number of time steps) was found to depend
roughly linearly on $\Dlt$, and for typical \Ref{poirier06bohmIII} values
($\Dlt=3.0$ a.u.,
$\tmax=6000$ a.u.) resulted in a final error of $10\%$.

\begin{figure}
\includegraphics[scale=0.7]{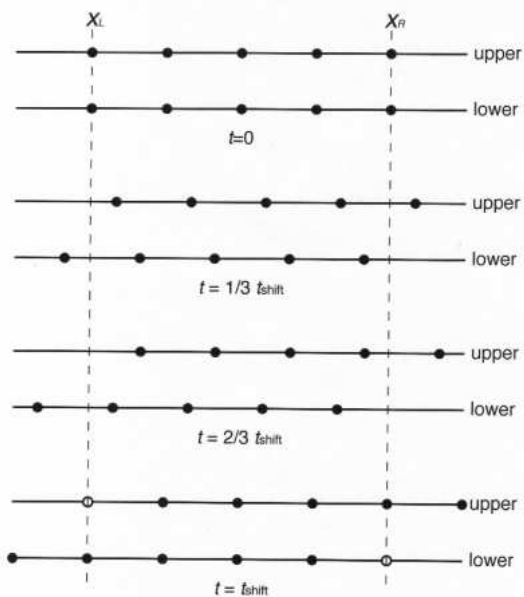}
        \caption{Schematic indicating location of trajectory grid points (circles)
                 over time, for upper ($\Psi_+$) and lower ($\Psi_-$) grids of the constant
                 velocity trajectory bipolar counterpropagating wave method. At
                 $t=0$, the $N=5$ points of both grids are distributed
                 uniformly from $x_L$ to $x_R$, with grid spacing $\Dlt x = (x_R-x_L)/(N-1)$.
                 Over time, the upper/lower grid points move with constant velocity
                 $\pm v=\pm \sqrt{ 2 E/m}$. At time $t_{\text{shift}} = \Dlt x /v $,
                 the two grids are again coincident in the interior region. The two
                 outermost points are deleted, and replaced with two new points
                 (indicated with open circles) at the opposite ends of their respective
                 grids, to restore the situation at $t=0$.}
        \label{gridfig}
\end{figure}

An effective solution to the asymptotic growth problem has been found, but
first it is necessary to divulge further details of the trajectory
implementation, as presented schematically in Fig.~\ref{gridfig}. At $t=0$,
the two trajectory grids, i.e. ``upper'' (for $\Pp$) and ``lower'' (for $\Pm$)
are coincident, with grid points at both $x_L$ and $x_R$, and a uniform grid
spacing of $\Dlt x = (x_R-x_L)/(N-1)$. Over time, $x^+_N$, the right-most
trajectory of the upper grid, evolves beyond the interaction region, as does
$x^-_1$, the left-most point of the lower grid. At time
$\tshift = \Dlt x / v$,
when these extremal points have progressed one grid spacing beyond
the interaction region, they are discarded, and new upper/lower grid
trajectories are introduced at $x_L$/$x_R$, and assigned $\Ppm$ values in
accord with \eq{bcs}. The indexing for both grids is shifted by one,
resulting once again in coincident grids identical to that at $t=0$. This approach
obviates the requirement of a nearest-neighbor search algorithm, thus greatly
reducing CPU cost per time step. It does require that $\tshift$ be a multiple of
$\Dlt$, although in practice, this poses a very minor constraint.

Note from Fig.~\ref{gridfig} that at general times within the cycle described
above, one or two points from each grid lie outside the range of the opposite
grid. In such cases, one might consider resorting to extrapolation to evaluate
the coupling term in \eq{Pdotunigwf}, although this approach is {\em not}
recommended.\cite{press} Instead, since the exterior points all lie
near the edges of the interaction region, we simply set $V\ra 0$ in
\eq{Pdotunigwf}. In principle, this results in unitary plane wave propagation
(PWP) for the exterior points. In practice, the Euler time integrator used
here is {\em not} unitary (nor is Runge-Kutta)---thus causing the observed growth
of asymptotic probability. This situation is well understood in the literature,
and various alternative unitary time integrators exist.\cite{press}
However, since the analytical plane wave solution is known, we simply adopt
the following exact unitary PWP scheme:
\eb
\Psi_{\pm k}(t+\Dlt) = \Psi_{\pm k}(t) \exp\sof{{i\over \hbar}  E \Dlt},
\label{PWP}
\ee
where $1 \le k \le N$ indexes individual trajectories.

The above PWP modification, applied to extreme points on {\em both} sides
of the two grids in lieu of Euler time integration, completely eradicates the
asymptotic growth of probability over time. However, it leads to a new problem,
i.e. asymptotic ``ramping.'' The idea here is that as the $x^+_1$ trajectory
travels inward, calculation of the plane wave portion of \eq{Pdotunigwf}
shifts suddenly from the unitary PWP to the nonunitary Euler integrator, causing the
final converged solution to manifest a gradual increase in asymptotic
$\rho_+(x)$ from left to right. This spurious ramping was found to be on the
order of a few percent for the test Eckart system---though the effect is once
again greatly exaggerated here, due to the first-order time integrator used.
Still, even with realistic time integrators (e.g. fourth-order Runge Kutta)
the ramping problem would cause difficulties for very high-accuracy
calculations. A simple solution is nevertheless available:
use the \eq{PWP} PWP for the plane wave portion of \eq{Pdotunigwf}
{\em throughout} the coordinate range (i.e. not just in the asymptotic
extremes). The first-order forward implementation is therefore as follows:
\begin{eqnarray}
\Psi_{\pm k}(t+\Dlt) && = \Psi_{\pm k}(t) \exp\sof{{i\over\hbar} E \Dlt}\nonumber\\
&&- {i\over\hbar} V(x^\pm_k)\Dlt \sof{\Psi_{\pm k}(t) + \Psi_{\mp}(x^\pm_k,t)},
\label{totalP}
\end{eqnarray}
where $\Psi_{\mp}(x^\pm_k,t)$ is computed via interpolation. Use of \eq{totalP}
has eradicated the nonunitary error completely in the test Eckart application
with parameters described above, resulting in a numerical solution accurate
to several parts per thousand throughout the full $x$ range. The error that
remains adheres to the form of interpolation error, as discussed above.

\subsection{Interpolation}
\label{interpolation}

As interpolation is a primary source of numerical error, it is essential
for the present purpose that this operation be performed in as efficient
and accurate a manner as possible. One of the key presumptions of successful
interpolation is that the underlying mathematical function be smooth and
well-behaved.\cite{wyatt,press} This results not only in more accurate
computation, but also better characterization of the numerical analysis.
Fortunately, the bipolar CPWM methodology is designed to provide smooth,
slowly-varying field functions even when $\Psi$ itself is not;
these have indeed been observed in all cases considered thus far,
with the F constant-velocity functions being particularly slowly-varying.
Other desired attributes of the interpolation routine include generality,
robustness, systematic error reduction, and a
minimum of required user intervention.\cite{wyatt}

In the previous work,\cite{poirier06bohmIII} fourth-order local polynomial
interpolation was used, analogous to a moving least squares method with a
stencil of five grid points for the interior of the grid.\cite{wyatt,press}
At the grid edges, fewer stencil points (and lower-order polynomials)
were used. In principle, if the field functions are sufficiently smooth, and
the grid points ideally distributed, then higher-order calculations can lead
to more accurate results. However, this is not guaranteed, and in some cases
higher-order interpolation may introduce unwanted wild oscillation between
grid points.\cite{press} As robustness and ease-of-use are key for the present
purpose as described above, we simply adopt third-order interpolation as the
most reasonable (and standard) choice. Also, since edge effects appear to be
so important (Sec.~\ref{oldanalysis}), we retain third-order interpolation
to the edges of the grids, i.e. four-point stencils are used
throughout for polynomial interpolation. We also consider spline
interpolation,\cite{press} for which the ``stencil'' is always two grid
points, in effect, regardless of order. Spline interpolation has the advantage
of continuous second derivatives throughout the coordinate range, yielding
``stiffer'' interpolants that are more stable, and generally more accurate
for smooth functions.

The numerical implementation described in Sec.~\ref{constant} employs
interpolation only when evaluating the coupling term,
i.e. the very last $\Psi_{\mp}(x^\pm_k,t)$ quantity in \eq{totalP}.
Although \eqs{PWP}{totalP} are written in terms of $\Ppm$, in fact it is the
real-valued field quantities, $r_\pm$ and $s_\pm$, obtained from
\eb
     \Ppm(x) = r_\pm (x) e^{i s_\pm (x)/\hbar}, \label{twoLM}
\ee
that are designed to be most smooth and slowly-varying
(especially in the classical
limit).\cite{poirier04bohmI,poirier06bohmII,poirier06bohmIII}
Numerically, one might consider storing $r_{\pm k}$ and $s_{\pm k}$ values
instead of $\Psi_{\pm k}$'s, but the corresponding \eq{totalP} becomes
complicated, nonlinear, and numerically less stable.

Instead, we adopt a ``best of both worlds'' approach, whereby $\Psi_{\pm k}$
values are stored and used in \eqs{PWP}{totalP}, $r_{\mp k}$'s and $s_{\mp k}$'s
obtained at each time-step, and the latter interpolated to compute the coupling
term in \eq{totalP} via \eq{twoLM}. Obtaining
$r_{\mp k} =  |\Psi_{\mp k}|$ is straightforward;
however, $s_{\mp k} = \hbar \arctan\of{\im \Psi_{\mp k}/ \re \Psi_{\mp k}}$ requires
special care in the choice of branch cut, to ensure continuity of the field
function throughout the coordinate range. Our simple solution to this well-known
dilemma\cite{wyatt} is the recursive implementation
\begin{widetext}
\eb
     s_{\mp (k+1)} =  \cases{s_{\mp k} & for $\Psi_{\mp (k+1)}=0$; \cr
        \hbar \arctan\!\sof{{\im \Psi_{\mp (k+1)}\over \re \Psi_{\mp (k+1)}} } &
             for $\Psi_{\mp k}=0$; \cr
        s_{\mp k} + \hbar \arctan\!\sof{{\im (\Psi_{\mp (k+1)}/\Psi_{\mp k})
               \over \re (\Psi_{\mp (k+1)}/\Psi_{\mp k})}} & otherwise, \cr }
\ee
\end{widetext}
with $\Psi_{\mp 0} = 0$ by definition. Note also that due to the regular
structure of the grid, there is no need for a search algorithm to compute
the stencil for a given interpolant point, i.e. the total CPU effort per
time-step associated with interpolation scales as $N$ rather than $N \log N$,
placing it on a par with potential function evaluation.

The above $r/s$ interpolation scheme yields phenomenal improvements in
accuracy when applied to the test Eckart problem of Sec.~\ref{constant}, as
compared to the $\Ppm$ interpolation applied there. The numerical solutions still
exhibit asymptotic oscillations; however, these are no longer constant in amplitude,
but damp exponentially with increasing $|x|$, as do the true mathematical
solutions. In any case, at the $-x_L = x_R = 3.0$ a.u. grid-edge
values considered, the amplitude has been reduced from a few parts per thousand
(Sec.~\ref{constant}) to a few parts per {\em million}, using a grid of only
$N=31$ points.

We conclude this subsection with a brief numerical comparison between
polynomial and spline interpolation (both third order) in the context of the
above algorithm as applied to the Eckart B problem with $E = 0.40 V_0$
(Sec.~\ref{eckart}). First though, as any complete spline implementation
requires specification of the boundary conditions, we
compared the accuracy of zero-second-derivative boundaries (natural
splines) vs. the more usual zero-first-derivative case, for known
functions similar to those computed below. The natural spline case was found to be
about an order of magnitude more accurate throughout the coordinate range,
and is therefore adopted.

For comparative purposes, both natural spline and polynomial interpolation
were applied to the Eckart B problem. Both cases are characterized by
quick convergence with respect to grid size, $N$, with the spline case
converging more quickly.
With respect to the time step, $\Dlt$, polynomial interpolation converges
more quickly at first---i.e., for larger $\Dlt$ values and lower accuracies.
Beyond a certain point however, the polynomial error flattens out and
even {\em increases} with subsequent decrease in $\Dlt$ (even if $N$ is
allowed to increase), whereas the spline error becomes arbitrarily small.
Further analysis reveals that this behavior of the polynomial interpolation
is due to asympotic oscillations, which increase in amplitude with decreasing
$\Dlt$, whereas in the spline case, the oscillation amplitudes are independent
of $\Dlt$. The polynomial case thus exhibits an effective minimum achievable
error, on the order of $10^{-5}$ for the case considered (note: in conjunction
with first-order forward time evolution). The parameters used
for this optimized polynomial calculation were also applied to a spline
calculation, which was found to be both more accurate (orders of magnitude
so for the transmission probability) and faster (by about 30\%). Taking these
facts into consideration, we adopt natural cubic spline interpolation from
here on out.

\subsection{Discontinuous Constant Velocity Scheme}
\label{discontinuous}

The constant-velocity trajectory method as discussed above is
not applicable for asymptotically asymmetric
potentials\cite{poirier06bohmIII}---i.e., $V_L = 0 \ne V_R$, where
$V_L = \lim_{x\ra -\infty} V(x)$ and
$V_R = \lim_{x\ra \infty} V(x)$. Mathematically, this is because the
coupling term in \eq{Pdotunigwf} does not vanish in the product asymptote
($x\ra\infty$), and so the $\Ppm(x,t)$ components do not converge over time.
Physically, it is because the trajectory velocity $v$ is not commensurate
with the product translational kinetic energy, $(E-V_R)$.
In any event, the presumption $V_L=V_R$ is too restrictive in practice to
be of general use, and so some appropriate resolution must be found.
One might consider the classical trajectory bipolar CPWM, for instance,
which would work fine for barrierless reactions; however, the more general
and realistic case of an asymmetric reaction {\em with} a barrier
(Fig.~\ref{asympotfig})---and therefore below-barrier energies---would seem
to be beyond the reach of either method.

On the other hand, the above difficulty is simply a manifestation of a standard
concern that is not at all new in reactive scattering theory---i.e., that the
product and reactant asymptotic potentials, and associated quantum states, are
not the same.\cite{taylor} Various remedies have been prescribed for this,
including use of a ``dividing surface''\cite{seideman92a,seideman92b} that partitions
configuration space into reactant and product sides. For 1 DOF applications,
the dividing ``surface'' is just a single point, labelled $x_0$. We can easily
incorporate the dividing point idea into the constant velocity scheme
by using trajectories with reactant asymptotic velocities in the $x<x_0$
region, and another set with product asymptotic velocities in the $x>x_0$ region.
The appropriate generalization of \eq{Pdotunigwf} (which holds for any
asymptotic values of the potential) is
\begin{eqnarray}
     {d \Psi_{L/R\pm} \over dt} &=& {i \over \hbar} \of{E-V-V_{L/R}}\Psi_{L/R\pm}\nonumber\\
          && -{i \over \hbar} \of{V-V_{L/R}} \Psi_{L/R\mp} \label{Pdotunigwfgen},
\end{eqnarray}
with ``$L$'' referring to the reactant (left) side of $x_0$, and ``$R$'' the
product (right) side. The corresponding trajectory velocities are
$v_{L/R} = \sqrt{2(E-V_{L/R})/m}$.

Although $\Psi$ itself, and its first spatial derivative, must be continuous
across $x=x_0$ at all times, the components $\Ppm$ are decidedly
{\em discontinuous}. In a trajectory implementation, $\Psi_{L+}(x_0)$ and
$\Psi_{R-}(x_0)$ are known quantities, and $\Psi_{L-}(x_0)$ and $\Psi_{R+}(x_0)$
are unknown; however, the $\Psi$ constraint above can be used to uniquely
specify the latter quantities in terms of the former.\cite{poirier06bohmIII}
For the sake of numerical efficiency, it is
advisable that $\Psi_{L+}$ and $\Psi_{R-}$ trajectories cross $x_0$ at
exactly the same time; when this occurs, these ``edge'' trajectories (for
their respective regions) are destroyed, and new edge trajectories
for $\Psi_{L-}$ and $\Psi_{R+}$ simultaneously created without needing to resort
to extrapolation. This is also when trajectory ``shifting'' occurs, as
per Sec.~\ref{constant}. The above trajectory crossing condition requires
that the grid spacings be different on either side of $x_0$, i.e.
$\Dlt x_{L/R} =  \tshift v_{L/R} $. In other respects, the
algorithm is similar to Sec.~\ref{constant}---except that when $x_k\ne x_0$,
we can not use \eq{PWP} for the edge trajectories near $x_0$, because
the coupling in this region is significant. Instead, we must use \eq{totalP},
even though this implies extrapolation within a range $(\Dlt x_{L/R})/2$ of $x_0$.

\subsection{Ramp Trajectory Scheme}
\label{general}

It would be beneficial to develop a single, robust, and efficient bipolar CPWM
scheme with none of the limitations of the previously-described methods. The method
should in other words be continuous, slowly-varying and extrapolation-free,
yet applicable to completely general potentials such as that of Fig.~\ref{asympotfig}.
From the discussion in Sec.~\ref{oldmethods}, it is natural to consider the
generalized F approach in this regard. Specifically, the ``semiclassical''
trajectories considered need have no direct relation to the actual potential
$V(x)$, but rather, are defined for some arbitrary effective potential,
$\Veff(x)$, as $v(x) = \sqrt{2\sof{E-\Veff(x)}/m}$. The specific
choice $\Veff(x) = V(x)$ reproduces the classical trajectory scheme,
$\Veff(x) = 0$ the constant-velocity scheme, and
$\Veff(x) = V_L + (V_R-V_L)\Theta(x-x_0)$ the discontinuous scheme.

The ideal $\Veff(x)$ should:
\begin{enumerate}
\item approach the flat $V(x)$ function in both asymptotic limits, i.e.
$\lim_{x\ra \mp\infty}\Veff(x) = V_{L/R}$
\item vary smoothly and monotonically through the interaction region.
\end{enumerate}
Condition 1. results in vanishing asymptotic coupling, whereas 2.
ensures there are no turning points, so that barrier tunneling poses
no difficulties. A sigmoid ramp function is thus clearly indicated;
below, we present a specific form based on the tanh function that is
applicable to extremely general situations. Although this form
might not be ideally suited to systems with reaction intermediates,
one could in that case add several ramp potentials together to
match the reaction profile appropriately. In any event, we must first work
out the theory and numerical implementation for the generalized F-based
CPWM approach.

The generalized F decomposition of \eq{psitot} for stationary scattering
states in 1 DOF is uniquely determined from the time-independent
\shro\ equation and the relation\cite{poirier06bohmIII,froman}
\eb
     \Psi' = -{v' \over 2v} \Psi +
          {i m v \over  \hbar} \of{\Psi_+ - \Psi_-}. \label{FPprimeB}
\ee
For completely general $v(x)$, these lead to
\eb
     \Ppm' = \sof{-{1\over 2}\of{{v' \over v}} \pm {i \over \hbar} m v
                     \pm {i \over 2} A } \Ppm
                     \pm {i \over 2} A \Pmp, \label{generalPpm}
\ee 
where
\eb
 A = \of{\hbar \over m v} \sof{{3 \over 4} \of{v'\over v}^2
                 - {1\over 2} \of {{v''\over v}} - {2 m \over \hbar^2}(V-\Veff)}.
     \label{Aex}
\ee

For consistency, it is convenient to reexpress the $v$ derivatives
above in terms of $\Veff$ and its derivatives. If the trajectory equations
of motion are to have no $\Ppm$ spatial derivatives as desired,
then we must construct a convective term somehow; this is
easily accomplished by multiplying \eq{generalPpm} by $\pm v$. The result is:
\ea{
     \pm v \Ppm' & = & \sof{ \pm {1\over 4} v \of{{\Veff' \over E-\Veff}}
                          + {i \over \hbar} \of{m v^2 - V + \Veff + C} }
                          \nonumber \\
                  & &  \times \Ppm - {i \over \hbar} \sof{V - \Veff - C} \Pmp,
                           \qquad \text{where} \label{convective} \\
      C & = & \of{\hbar^2 \over 2m} \sof{{5 \over 16} \of{{\Veff'\over E-\Veff}}^2 +
                                        {1 \over 4} \of{{\Veff''\over E-\Veff}} }. }
Next, we replace $m v^2$ in \eq{convective} with $2(E-\Veff)$.
Adding $-(i/\hbar) E \Ppm \mp v \Ppm'$ to both sides, and recognizing that for
the stationary solution, $\partial \Ppm /\partial t = -(i/\hbar) E \Ppm$,
we obtain finally
\ea {
     {\partial \Ppm \over \partial t} & = &\mp v \Ppm'\nonumber\\
&& + \sof{ \pm {1\over 4} v \of{{\Veff' \over E-\Veff}}
                          + {i \over \hbar} \of{E - V - \Veff + C} } \Ppm \nonumber \\
                  & &  \qquad\quad - {i \over \hbar} \sof{V - \Veff - C} \Pmp.
                           \label{Pdotgeneral} }

Note that for constant velocity trajectories, $C = \Veff = 0$, and
\eq{Pdotgeneral} reduces to \eq{Pdotuni}, as it should. Also interesting
(and useful) is the fact that \eq{Pdotgeneral} satisfies the combined
continuity relation\cite{poirier06bohmIII}---i.e. the total combined probability
($\rho_+ + \rho_-$) is conserved over time, even though the individual
$\rho_\pm$ are not. The precise flux relation giving rise to this is
\eb
     {\partial \rho_\pm \over \partial t} = - j_\pm' \pm
       {2 \over \hbar}\of{V-\Veff -C}
       \im \sof{{\Psi_+}^* \,\Psi_-}, \label{fluxrel}
\ee
where $j_\pm = \pm v \rho_\pm$ is the flux, defined in terms of the generalized
trajectory velocities. For the constant velocity case, this reduces to
\Ref{poirier06bohmIII} Eq.~(13).

In implementing the above scheme numerically, we borrow from the
successful constant-velocity algorithm as much as possible. The condition that
over a duration of time, $\tshift$,
each grid point should progress to the precise spot vacated
by its nearest neighbor, implies a very nonuniform grid---with a higher density
of points in areas where $v$ is lower, as is appropriate. Though by no means
necessary, we find it convenient to compute all of the trajectories in advance.
Actually, only the single trajectory $x^+_1(t)$ need be computed---starting at
$x=x_L$, and propagating classically in accord with $\Veff$ to a bit past
$x=x_R$. The other trajectories can then be obtained via
$x^+_k(t) = x^+_1\sof{t+(k-1)\tshift}$,
and $x^-_k(t) = x^+_k(-t)$. The $x^+_1(t)$ values for each positive integer
multiple of the CPWM time step, $\Dlt$, are computed and stored during this
preprocessing phase, as are the corresponding effective potential values
$\Veff$, $\Veff'$, and $\Veff''$---a technique that can in principle save much
CPU time during the subsequent CPWM propagation phase. Note that by design,
the trajectory version of \eq{Pdotgeneral} involves no explicit spatial
differentiation of $\Ppm$; once again, the largest source of numerical error
(and CPU effort) is interpolation. To ensure that
the trajectory calculation does not contribute appreciably to the
error, it is performed using a much smaller time step, $\delta = \Dlt/100$;
even so, the CPU cost of the preprocessing phase is a trivial part
of the whole cost. Also, numerical comparisons with $\delta = \Dlt/30$
and $\Dlt/300$ reveal no discernible difference in computed results.

Another difference from the constant-velocity scheme is the choice of
initial conditions. Based on the discussion in Sec.~\ref{oldmethods}, we
take $\Psi_+(t=0)$ to be the basic WKB solution, i.e.
\eb
    \Psi_+(x,0) = \sqrt{v_L/v(x)} \exp \sof{ i s_0(x)/\hbar},
\ee
where the initial action, $s_0(x)$, is also computed during the preprocessing phase
(the initial reflected wave is of course still zero).
The generalization of \eq{totalP} is seemingly non-trivial:
\ea{
     \Psi_{\pm k}(t+\Dlt) & = & \Psi_{\pm k}(t)\\
& & \exp\bof{{i\over\hbar} \Dlt
            \sof{E - 2 \Veff \pm {i\over 4} \hbar v \of{{\Veff'\over E-\Veff}} } }
                             \nonumber \\
                          & & - {i\over \hbar} \Dlt \of{V-\Veff -C}
                             \sof{\Psi_{\pm k}(t) + \Psi_{\mp}(x^\pm_k,t)},\nonumber
     \label{totalPgen} }
with all functions evaluated at $x^\pm_k$.
The rationale is the same as before however: the asymptotically
vanishing coupling term in \eq{Pdotgeneral} (but for $\Psi$, rather
than just $\Ppm$) is treated using forward Euler propagation;
everything else is exponentiated as per \eq{PWP}.

The proceeding discussion is completely general, i.e. applies to any choice
of $\Veff(x)$. However, for single-barrier reactions (no intermediates), we
advocate the following analytical ramp functional form, whose derivatives
are also specified:
\ea{
     \Veff(x) & = & \of{{V_R-V_L \over 2}} \bof{\tanh \sof{\beta (x-x_0)}+1}
     \label{uphillramp} \\
     \Veff'(x) & = & \of{{V_R-V_L \over 2}}
                              \bof{{\beta \over \cosh^2\sof{\beta (x-x_0)} }}
                     \nonumber \\
     \Veff''(x) & = & -(V_R-V_L) \beta^2 \bof{{ \tanh \sof{\beta (x-x_0)}
                               \over \cosh^2\sof{\beta (x-x_0)} }} \nonumber}
As in Sec.~\ref{discontinuous}, $x_0$ should be in the middle of the $V(x)$
interaction region, and $1/\beta$ comparable to the interaction region size.


\subsection{Summary and Numerical Recipes}
\label{summary}

To summarize, the key numerical developments in this section have been:
(1) use of trajectory grids that are coincident when $t$ is a multiple
$\tshift$ (implying that $\tshift$ itself is a multiple of $\Delta$);
(2) unitary plane wave (or exponential) propagation throughout the coordinate range;
(3) third-order spline interpolation of $r_\pm$ and $s_\pm$, to evaluate the
coupling contribution to the time-evolution equations.

For asymptotically symmetric potentials, the recipe for numerical propagation of
the constant velocity trajectory quantities is as follows:
\begin{itemize}
\item Move each grid point, $x_k^\pm$, via $x_k^\pm(t+\Delta) = x_k^\pm(t) \pm v \Delta$
\item For exterior grid points outside the range of the opposite grid, update
$\Psi_{\pm k}$ via \eq{PWP}.
\item For interior grid points, update $\Psi_{\pm k}$ via Eqs.~(\ref{totalP}),
(\ref{twoLM}), and (9).
\end{itemize}

For asymptotically asymmetric potentials, the ramp trajectory method is
preferred over the discontinuous trajectory method (Sec.~\ref{results}).
The ramp trajectories are computed and stored {\em a priori}, in
a preprocessing step. The subsequent numerical propagation is as follows:
\begin{itemize}
\item For exterior grid points outside the range of the opposite grid, update
$\Psi_{\pm k}$ via the first line of \eq{totalPgen}.
\item For interior grid points, update $\Psi_{\pm k}$ via Eqs.~(\ref{totalPgen}),
(\ref{twoLM}), and (9).
\end{itemize}


\section{RESULTS}
\label{results}

In this section, the three different numerical algorithms described
in detail in Sec.~\ref{theory}---i.e. the constant velocity, discontinuous
constant velocity, and ramp trajectory versions of the non-wave-front
F trajectory method---are applied to a variety of test applications,
in order to assess them for robustness and numerical efficiency.
In all cases, first-order forward Euler time-evolution is employed
as discussed previously, leading to ``artificially'' small time steps, $\Dlt$
(e.g. as compared to the fourth-order Runge-Kutta results of \Ref{poirier06bohmIII}).
The mass $m=2000$ a.u. is used in all applications.

A thorough and detailed convergence study was performed for each
application as follows. First, given the small oscillations with $x$
that characterize the numerical $\rho_\pm(x)$ solutions in the asymptotic
regions, the usual determination of reflection and transmission probabilities
as $P_{\text{refl}} =|\Psi_-(x_L)|^2$;
$P_{\text{trans}} = \of{v_R/v_L}|\Psi_+(x_R)|^2$
is not used. Instead, a small range of points near $x_{L/R}$ is consulted
to characterize both mean values {\em and uncertainties} for these quantities.
Reducing the oscillatory uncertainty to an acceptably small level is
one of two conditions that must be satisfied to achieve convergence; the second is that
variation of the mean $P_{\text{refl}}$ and $P_{\text{trans}}$ values
with respect to all numerical parameters be acceptably small. The convergence
parameters, in the cyclical order in which they are varied (keeping all
others fixed) are as follows: maximum time, $\tmax$; grid spacing, $\Dlt x$;
time step, $\Dlt$; grid boundaries, $x_{L/R}$. For each of the applications
considered, tens of convergence calculations were performed, usually over
multiple cycles.

\subsection{The Eckart Barrier}
\label{eckart}

\begin{figure}
\includegraphics[scale=0.7]{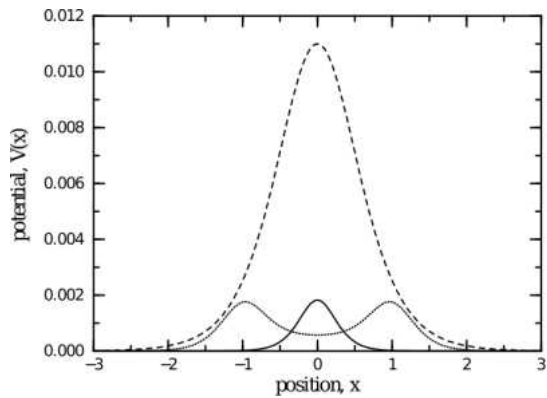}
        \caption{Potential energy curves for several symmetric potential systems considered
         in this paper: Eckart A barrier (solid); Eckart B barrier (dashed);
         double barrier (dotted). All units are atomic units.}
        \label{sympotfig}
\end{figure}

The first application considered is the symmetric
Eckart barrier,\cite{eckart30,ahmed93}
\eb
     V(x)=V_0\,\text{sech}(\alpha x)^2, \label{eckartpot}
\ee
with parameter values $V_0=400\, \text{cm}^{-1}$,
and $\alpha=3.0$ a.u, which for the present paper is
labeled ``Eckart A.'' This application is identical to the
Eckart problem of \Ref{poirier06bohmIII}, and is chosen here in
part to compare with the previous results. A plot of
this potential is presented in Fig.~\ref{sympotfig}.

A detailed convergence study was performed for the energy equal to
the barrier peak, i.e.
$E = V_0 = 400\, \text{cm}^{-1} \approx 0.0018$ hartree.
This energy was chosen because it provides a good splitting of
probability between $P_{\text{refl}}$ and $P_{\text{trans}}$.
Of the three numerical methods considered, only the
constant velocity method is applicable here---because
both the discontinuous constant velocity and ramp trajectory
methods reduce to the constant velocity method
when the potential $V(x)$ is symmetric.
In any event, the parameters that resulted from the convergence
procedure described above, with $10^{-4}$ as the target accuracy for
both $P_{\text{refl}}$ and $P_{\text{trans}}$, are presented in
Table~\ref{sympottab} column~3. These can be compared directly
to the \Ref{poirier06bohmIII} calculation (computed to the same level of accuracy),
presented in column~2, as well as to the analytical values given in
the last two rows.

\begin{table*}
\caption{Convergence parameters for constant velocity trajectory method
applied to several symmetric potential systems, with energies near
or at the barrier height. All units are atomic units.
Column 2: older algorithm used in Ref.~3.
Columns 3--6: newer algorithm of Sec.~\ref{constant}. Convergence of
quantities indicated is to $10^{-4}$ for Eckart A calculations, and $10^{-3}$
for others. Rows 7 and 8: computed reflection and transmission
probabilities, including oscillatory uncertainties. Last two rows: exact
reflection and transmission probabilities for Eckart systems.}
\label{sympottab}
\begin{ruledtabular}
\begin{tabular}{cccccc}
Quantity & \multicolumn{5}{c}{Symmetric potential system} \\ \cline{2-6}
   and      & \multicolumn{3}{c}{Eckart A} & Eckart B & double Gaussian
              \\ \cline{2-4}
 symbol     & \Ref{poirier06bohmIII} & $P_{\text{refl}}$ and $P_{\text{trans}}$ &
           $P_{\text{refl}}$ only & $P_{\text{refl}}$ only &
           $P_{\text{refl}}$ and $P_{\text{trans}}$ \\ \hline
grid size,       $N$                & 31    & 20    &  13   & 25    & 20    \\
left edge,       $x_L$              & -3.0  & -2.0  & -2.0  & -3.0  & -3.0  \\
right edge,      $x_R$              &  3.0  &  2.0  &  1.5  & 2.1   & 2.5   \\
grid spacing,    $\Dlt x$           & 0.20  & 0.211 & 0.292 & 0.213 & 0.289 \\
time step,       $\Dlt$             & 10.0  & 0.156 & 0.324 & 0.046 & 0.612 \\
max time,        $t_{\text{max}}$   & 10000 & 3899  & 4105  & 3204  & 41015 \\
computed         $P_{\text{refl}}$  &       & $.28336\pm.00004$ & $.28323\pm.00010$ &
                                              $.4587 \pm .0002$ & $.7936 \pm .0002$ \\
computed         $P_{\text{trans}}$ &       & $.71646\pm.00005$ & $.71625\pm.00005$ &
                                              $.5355 \pm .0003$ & $.20498\pm.00001$ \\
exact            $P_{\text{refl}}$  &       & 0.283358 & 0.283358 & 0.459605 & \\
exact            $P_{\text{trans}}$ &       & 0.716642 & 0.716642 & 0.540395 & \\
\end{tabular}
\end{ruledtabular}
\end{table*}

The table indicates that the new algorithm requires substantially
fewer grid points than the old---a testament to the efficacy of the
new PWP and interpolation routines. However, the new algorithm
requires a {\em much } smaller time step (more than $50\times$),
owing to the fact that the non-PWP time integrator is only first-order,
rather than fourth-order. Presumably for the same reason, $\Delta$ was
found to be the slowest parameter to converge. From
Table~\ref{sympottab}, it is also clear that the computed $P_{\text{refl}}$
value is substantially more accurate than $P_{\text{trans}}$. Since only
one of these two quantities is needed in practice (because their sum is unity),
we performed a second convergence study for which $P_{\text{refl}}$ alone was
converged to $10^{-4}$. The results, presented in Table~\ref{sympottab} column~4,
indicate a very substantial efficiency improvement, requiring a mere
$N=13$ grid points and 0.88 seconds on a 2.60 GHz Pentium CPU---as compared
to $N=31$ and 5.0 seconds for column~2. Note that for both calculations, the
oscillatory error is substantially smaller than the overall convergence
error---i.e. it would have been essentially just as accurate to use
$P_{\text{trans}} = \rho_+(x_R)$, etc.

Finally, column~5 of Table~\ref{sympottab} presents results for the
much broader and taller ``Eckart B'' barrier, i.e. with parameter values
$V_0=0.011$ hartree ($\approx 2414\, \text{cm}^{-1}$), and $\alpha=1.364$ a.u.
The Eckart B potential is also presented in Fig.~\ref{sympotfig}.
The goal here is to determine whether the new algorithm is robust
across a wide range of system parameters, and to what extent the
much larger-action Eckart B system requires additional computational
effort. The Eckart B system is also used to analyze
deep tunneling performance in Sec.~\ref{deeptunnel}. Here, though,
we consider only the barrier energy $E = V_0$.
A comparison between columns~4 and~5 reveals that a larger grid and
smaller time step are needed in the Eckart B case, as would be
expected based on system parameters. However, the near
order-of-magnitude time step reduction is far greater than
expected---particular given that the Eckart B calculation is
less accurate to one digit of precision. This is due to the convergence
of computed error with decreasing $\Dlt$, which, though initially fast,
is found to become quite slow (essentially inverse scaling) beyond a
certain threshold accuracy level. For Eckart B, this threshold appears
to be around $3 \times 10^{-3}$. In any case, higher-order time evolution
is expected to resolve this difficulty completely.

\subsection{The Deep Tunneling Regime}
\label{deeptunnel}

To analyze the efficacy of the constant velocity method in the deep
tunneling regime, the Eckart B problem was solved for a range of
below-barrier energies, $E<V_0$. The large size of this barrier
ensures that even energies relatively close to the top of the
barrier---e.g., $E = 0.8\, V_0$---are characterized by quite
small transmission probabilities, $P_{\text{trans}}$. By
$E = 0.1\, V_0$, the $P_{\text{trans}}$ values are extremely small
(on the order of $10^{-9}$). Detailed convergence studies were
performed for the above two energy values, as well as for the
intermediate value $E = 0.4 \,V_0$,
in order to assess the method across the whole energetic range.
The details are presented in Table~\ref{deeptab}.
For simplicity, $x_L = x_R$ was presumed. The quantity $P_{\text{trans}}$
was converged to an absolute accuracy that varied substantially
with $E$, as indicated in Table~\ref{deeptab} row 6. In relative
accuracy terms, this translates to three-to-five significant digits
for each computed $P_{\text{trans}}$ value, with the $E = 0.4\, V_0$
case substantially more accurate than the other two. In all cases, the
convergence error is comparable to the actual error, obtained
relative to the exact $P_{\text{trans}}$ values presented in row 8.
These findings are particularly satisfying for the extreme $E=0.1\, V_0$
case, which is computed here to an absolute error of $10^{-12}$ or so.

\begin{table*}
\caption{Convergence parameters for constant velocity trajectory method
applied to Eckart B system over a wide range of below-barrier energies,
$E<V_0$. All units are atomic units. Convergence is of transmission
probability, $P_{\text{trans}}$, only, to accuracy level indicated in
row 6. Computed $P_{\text{trans}}$ values including oscillatory uncertainty
are listed in row 7, with exact values given in row 8.}
\label{deeptab}
\begin{ruledtabular}
\begin{tabular}{cccc}
Quantity   & \multicolumn{3}{c}{Energy $E$ as fraction of barrier height $V_0$.}
              \\ \cline{2-4}
and symbol & $0.8\,V_0$ & $0.4\,V_0$ & $0.1\,V_0$ \\ \hline
grid size,       $N$                   & 33        & 88          & 133          \\
grid edges,      $x_{L/R}$             & $\mp3.0$  & $\mp4.0$    & $\mp4.0$     \\
grid spacing,    $\Dlt x$              & 0.188     & 0.092       & 0.061        \\
time step,       $\Dlt$                & 0.126     & 0.097       & 1.156        \\
max time,        $t_{\text{max}}$      & 4424      & 46424       & 462285       \\
target accuracy for $P_{\text{trans}}$ & 0.02(-2)  & 0.0005(-5)  & 0.01(-10)    \\
computed         $P_{\text{trans}}$    & 4.444$\pm$.020(-2) & 1.5588$\pm$.0004(-5) &
                                         9.920$\pm$.002(-10) \\
exact            $P_{\text{trans}}$    & 4.462(-2) & 1.5594(-5)  & 9.920(-10) \\
\end{tabular}
\end{ruledtabular}
\end{table*}

The convergence trends with respect to decreasing $E$, as evident
in Table~\ref{deeptab}, are illuminating. Perhaps most striking is the
fact that only a modest increase in coordinate range, $x_{L/R}$, is
observed---completely unlike the usual situation for conventional
quantum scattering methods (see below). A modest increase in grid size,
$N$, arising primarily from a reduction in grid spacing, $\Dlt x$,
is nevertheless required. This is true, even though the wavelength
increases with decreasing $E$, because the lower energy calculations must
be performed to a substantially higher level of absolute accuracy.
Comparing rows 1 and 6, the convergence appears to be exponentially
fast, with something like twelve extra points needed for each additional
digit of accuracy. Relative accuracy also plays a role in this regard,
however, e.g. the $0.8 \, V_0$ calculation done to the same {\em absolute}
accuracy as the $0.4 \, V_0$ calculation is found (not presented
in Table~\ref{deeptab}) to require a greater number of grid points
than the latter.

There is a very substantial growth of $t_{\text{max}}$ with decreasing
energy, owing to the fact that the trajectories move
much more slowly. However, for the same reason, there is
also a substantial increase in the time step size, $\Dlt$, so that the
growth in the total {\em number} of time steps (and CPU effort) is much
more modest. The above comments apply for calculations at comparable
{\em relative} accuracies, e.g. Table~\ref{deeptab} columns 2 and 4.
The intermediate calculation (column 3), done to a substantially
higher relative accuracy than the others, is found to be in the
regime of slow convergence with respect to $\Dlt$ (Sec.~\ref{eckart})
and therefore manifests a much smaller converged value for this parameter.

The extreme $E= 0.1\, V_0$ calculation represents a very challenging test
case for conventional quantum dynamics methods. For comparative purposes,
an optimized sinc-DVR calculation\cite{seideman92a,seideman92b,colbert92}
with quartic absorbing boundary conditions was also performed. Such
methods always increase the coordinate range beyond the interaction
region---to a degree proportional to the wavelength, and thus very substantial
at low energies. For instance, $x_{L/R} = \mp 40.0$ a.u. for
the DVR calculation, a full order of magnitude larger than for
the CPWM calculation. The resultant grid size, $N=700$, was also
much larger---particularly given that CPU effort scales as $N^3$ for
DVR methods, but only as $N$ for the present approach.

The above discrepancies are magnified even more for shorter, narrower
barriers than Eckart B. In this context, deep tunneling is observed
only when $E$ is a tiny fraction of $V_0$. For the Eckart A problem
with $E = 10^{-4}\, V_0$, for instance, attempts to perform a converged
sinc-DVR calculation proved unsuccessful, owing to the enormous
coordinate range required. Instead, an approximate, semiclassical
tunneling calculation was performed, giving rise to the prediction
$P_{\text{trans}}= 3.82\times 10^{-3}$. In fact, this is more than $100\times$
larger than the actual value, $P_{\text{trans}}= 2.851\times10^{-5}$.
These examples serve to underscore the difficulty of computing accurate
deep tunneling transmission probabilities using conventional methods,
which may be greatly alleviated with the present bipolar CPWM approach.

\subsection{The Uphill Ramp}
\label{uphill}

\begin{figure}
\includegraphics[scale=0.7]{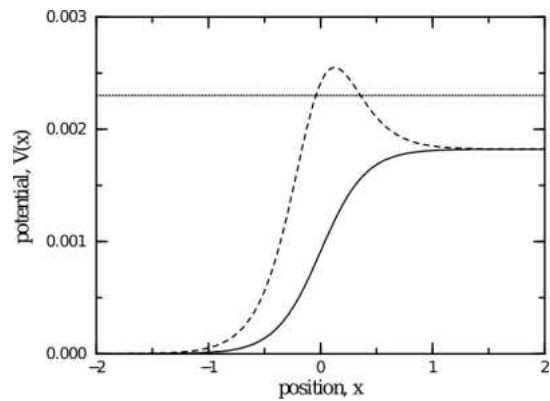}
        \caption{Potential energy curves for two asymmetric potential systems considered
         in this paper: uphill ramp (solid); barrier ramp (dashed). Dotted line indicates
         energy value, $E= 0.0023$, used for both systems. All units are atomic units.
         For the ramp trajectory calculations for {\em both} systems, the trajectory
         potential, $\Veff(x)$, is the solid curve.}
        \label{asympotfig}
\end{figure}

The first asympotically asymmetric system considered here
is the continuous ``uphill ramp,''\cite{flugge} defined via
\eq{uphillramp} and the parameter values
$V_L = 0$, $V_R = 400\, \text{cm}^{-1} \approx 0.0018$,
$x_0 = 0$, and $\beta=2.5$ a.u. Thus, $V(x) = \Veff(x)$,
so that the ramp trajectories are themselves classical
trajectories. The energy $E = 0.0023$ a.u
$\approx 500\, \text{cm}^{-1}$ is deliberately
chosen to be only moderately larger than $V_R$, to accentuate
the asymmetric difference between $v_L$ and $v_R$.
A plot is presented in Fig.~\ref{asympotfig}.
The system and energy parameters are identical to those considered
in \Ref{poirier06bohmIII}, and
are also comparable to those of the Eckart A calculation
in Sec.~\ref{eckart}, enabling a fairly direct comparison.
Detailed convergence studies were performed for both the
discontinuous constant velocity and ramp trajectory methods,
as presented in Table~\ref{asympottab}. Both $P_{\text{refl}}$
and $P_{\text{trans}}$ were converged, with the former to a
higher level of accuracy, appropriate to the different scales
of the two quantities. Note that this is the first example
considered for which $P_{\text{trans}} \gg P_{\text{refl}}$.

\begin{table*}
\caption{Convergence parameters for discontinuous constant velocity and
ramp trajectory methods, applied to two asymmetric potential systems
with energy $E = 0.0023$ a.u $\approx 500\, \text{cm}^{-1}$ slightly
larger than asymptotic potential difference $(V_R-V_L) =  400\, \text{cm}^{-1}$.
All units are atomic units. Columns 2 and 3: uphill ramp system. Columns
4 and 5 barrier ramp system. Convergence is of both $P_{\text{refl}}$ and
$P_{\text{trans}}$ to accuracy levels indicated in rows 7 and 8. Last two
rows: computed reflection and transmission probabilities, including oscillatory
uncertainties.}
\label{asympottab}
\begin{ruledtabular}
\begin{tabular}{ccccc}
Quantity & \multicolumn{4}{c}{Asymmetric potential system} \\ \cline{2-5}
   and      & \multicolumn{2}{c}{uphill ramp} & \multicolumn{2}{c}{barrier ramp}
              \\ \cline{2-3} \cline{4-5}
 symbol     & discontinuous & ramp traj. & discontinuous & ramp traj. \\ \hline
grid size,         $N$                 & 80       & 25         &  80       & 23     \\
left edge,         $x_L$               & -2.0     & -2.0       & -2.0      & -2.0   \\
right edge,        $x_R$               &  2.0     &  2.0       &  2.0      & 2.5    \\
left grid spacing, $\Dlt x_L$          & 0.080    & 0.252      & 0.080     & 0.315  \\
time step,         $\Dlt$              & 0.235    & 0.500      & 0.754     & 0.300  \\
max time,          $t_{\text{max}}$    & 6594     & 5478       & 7912      & 7474   \\
target accuracy for $P_{\text{refl}}$  & .00004   & .00002     & .0002     & .00008 \\
target accuracy for $P_{\text{trans}}$ & .0004    & .00015     & .0006     & .0004  \\
computed           $P_{\text{refl}}$   & $.023838 \pm .000002$ & $.023919 \pm .000003$ &
                                         $.45452 \pm .00003$   & $.45454 \pm .00003$ \\
computed           $P_{\text{trans}}$  & $.97557 \pm .00001$   & $.9762 \pm .0001$ &
                                         $.54465 \pm .00003$   & $.54602 \pm .0001$  \\
\end{tabular}
\end{ruledtabular}
\end{table*}

The discontinuous constant velocity case is considered first,
i.e. Table~\ref{asympottab} column 2. As compared with the
Eckart A calculation in Table~\ref{sympottab} column 3, the
most compelling difference is that a much higher density
of grid points is needed ($N=80$ vs. $N=20$), even though the
final computed accuracy is somewhat lower. This is most likely
attributable to the use of extrapolation (Sec.~\ref{discontinuous})
in the middle of the interaction region where the coupling is greatest.
One indication is the fact that the asymptotic oscillations
are much smaller than for Eckart A, and not a contributing factor
to the total error. Another significant difference is found in the
parameter $t_{\text{max}}$, which is larger in the uphill ramp case,
probably owing to the fact that product trajectories are
substantially slower than for Eckart A. Perhaps for the same reason,
$\Dlt$ is also somewhat larger for the uphill ramp.

The ramp trajectory convergence parameters are shown in
Table~\ref{asympottab} column 3. The $N=25$ grid size is now greatly
reduced in comparison to column 2, almost to the Eckart A level,
again suggesting that the primary source of error in the discontinuous
case is extrapolation. The small increase in $N$ relative to Eckart A
may be due to the fact that grid spacing in product and reactant
regions are not independently adjustable (which may also be a
contributing factor in the discontinuous calculation). Another key difference
between the two asymmetric calculations is that $t_{\text{max}}$ is
substantially reduced in the ramp trajectory case---most likely
reflecting the fact that the initial semiclassical approximation
$\Psi_+(x,0)$ is much closer to the actual solution (Sec.~\ref{general}).
Nevertheless, the time step $\Dlt$ is substantially larger. Taken together,
all of the above findings imply that CPU effort for the ramp trajectory
calculation is greatly reduced in comparison to the discontinous calculation.

\subsection{The Barrier Ramp}
\label{barrier}

The barrier ramp system (Fig.~\ref{asympotfig}) is the first asymmetric
reaction with a barrier to be treated using CPWM methods. Serving as
representative of a generic reaction profile, moreover, it is an extremely
important benchmark system. This is especially true for the ramp
trajectory method, because $\Veff(x)$, being monotonic, can no longer be
equal to $V(x)$, which might in principle affect computational efficiency.
To construct a suitable barrier ramp potential, we simply add an Eckart barrier
and an uphill ramp together. The Eckart parameters [\eq{eckartpot}] are
taken to be $V_0=0.0015$ a.u. and $\alpha=2.5 $ a.u, whereas the uphill
ramp is identical to that of Sec.~\ref{uphill}. The ramp trajectories
are also identical to Sec.~\ref{uphill}, i.e. $\Veff(x)$ is the same as
before. The energy $E = 0.0023$ a.u is also the same as in Sec. ~\ref{uphill},
as are the asymptotic trajectory velocities, $v_{L/R}$.

The discontinuous and ramp trajectory convergence parameters are shown in
Table~\ref{asympottab} columns 4 and 5, respectively. In both cases the
grid parameters are nearly identical to the corresponding uphill ramp
calculations (columns 2 and 3)---implying that the presence of the barrier
introduces no numerical difficulties for either method. The $t_{\text{max}}$
values are somewhat larger, undoubtedly owing to the fact that the barrier
ramp case has much greater reflection.\cite{poirier06bohmII,poirier06bohmIII}
For the discontinuous calculation, $\Dlt$ is lower than the uphill ramp value,
perhaps because the absolute accuracy of the computed $P_{\text{refl}}$ is
much less. This situation is reversed for the ramp trajectory calculation,
for which a somewhat smaller $\Dlt$ is required. The reason may have to do
with the fact that $\Veff(x) \ne V(x)$, giving rise to larger coupling
in \eq{Pdotgeneral}.

\subsection{The Double Barrier}
\label{double}

The last case to be considered is a system with reaction intermediates.
In \Ref{poirier06bohmIII}, a double-Gaussian barrier was employed. Here
however, we wish to have more flexibility in the functional form, in order
that a wider range of reaction profiles may be considered. To achieve this,
a sum of two barrier ramps is used, centered at the transition
states on either side of the reaction intermediate. Though symmetry
is not required, for simplicity we adopt the symmetric form here as follows:
\ea{
     V(x) & = & V_0\,\bof{\text{sech}[\alpha (x+x_0)]^2 +
                          \text{sech}[\alpha (x-x_0)]^2} + \nonumber \\
          & & (V_\Delta/2) \bof{\tanh\sof{\beta(x+x_0)} - \tanh\sof{\beta(x-x_0)}}\nonumber\\
& &  }
The parameters $V_0=0.0015$ a.u, $x_0 = 1.0$, $\alpha=\beta= 2.5$ a.u., and
$V_\Delta = 0.0005$ a.u. are comparable to those used for the barrier ramp and
Eckart A systems. The energy, $E = 0.0014$ a.u, lies slightly
below the barrier peak of around $0.00175$ a.u. A plot is presented in
Fig.~\ref{sympotfig}.

Although this system would be a good candidate for a double ramp trajectory
scheme (Sec.~\ref{general}), the constant velocity trajectory algorithm is used
here. Convergence for both $P_{\text{refl}}$ and $P_{\text{trans}}$ to
$10^{-3}$ is indicated in Table~\ref{sympottab} column 6. This accuracy is somewhat
lower than for Eckart A (column 3), to avoid slow $\Dlt$ convergence
(Sec.~\ref{eckart}). The coordinate range is larger than for Eckart A, but the
grid density is actually somewhat {\em smaller} (despite the fact that the potential
has more features), such that the $N=20$ grid size is the same for both.
The convergence time $t_{\text{max}}$ is much larger than for Eckart A, as
is expected for a system with tunneling and reaction
intermediates.\cite{poirier06bohmII,poirier06bohmIII}
In other respects, the calculation is similar to, and no more difficult than,
the other calculations without reaction intermediates.

\section{CONCLUSION}
\label{conclusion}

Based on the results as presented above, we conclude that
all of the primary goals of this paper have been achieved.
In particular, the robustness of the methods are demonstrated
by the very broad range of test cases considered.
These may be definitively regarded as ``representative'' of
many other 1 DOF calculations (not described here)
that were also performed. The efficiency of the methods is
also established, in that most of the calculations that do not
involve deep tunneling or reaction intermediates require no
more than about one second of CPU time. Finally, the accuracy
levels that can be achieved have been greatly expanded beyond
previous limits, e.g., to twelve-to-thirteen digits past the
decimal in the extreme deep tunneling case
(Table~\ref{deeptab} column 4)---probably the highest
accuracy ever achieved in a trajectory-based quantum calculation.
Although 1 DOF systems are usually regarded as trivially soluble,
these can nevertheless pose severe numerical difficulties when
extremely deep tunneling is involved; it is therefore
particularly encouraging to see that the new methods work
well in both this limit, and for more typical molecular situations.

Indeed, there do not appear to be any 1 DOF applications for
which the present algorithms would not work well. This is
of the utmost importance for the main motivation of the
present work, i.e. creating a parameter-free (except for convergence
parameters) ``black-box'' quantum dynamics code for solving
any 1 DOF stationary scattering problem.  A single universal
algorithm is sought, although three specific implementations were
considered in this paper: (1) constant velocity ;
(2) discontinuous constant velocity; (3) ramp trajectory. As
both (2) and (3) reduce to (1) in the asymptotically symmetric
potential case, the only real choice is between (2) and (3). For
reasons elucidated in Sec.~\ref{results}, (3) is clearly the
better choice, both with respect to numerical efficiency
{\em and} robustness. As evidence of the latter, about the same number of
grid points is required for systems with similar parameters,
regardless of whether the potential is asymmetric, or has one
or more barriers (compare Table~\ref{sympottab} columns~3 and 6
with Table~\ref{asympottab} columns 3 and 5).

The ramp trajectory scheme, (3), is therefore proposed for the universal
purpose described above. One can imagine a straightforward procedure
to automatize the selection of suitable ramp parameters for any
given potential without reaction intermediates. Such a procedure
would presumably be easily generalized to multiple ramp trajectories
in the case of reaction intermediates, thus maintaining
``black-boxness'' for all cases. On the other hand, even a single ramp
is seen to perform very well for multiple barrier systems
[as in Sec.~\ref{double}, due perhaps to the smoothness of $\Veff(x)$],
so that a single ramp may be all that is ever actually needed in practice.
In fact an earlier preliminary investigation (without the numerical refinements
of Sec.~\ref{theory}) obtained slightly faster convergence and smoother
field functions for single vs. double ramp calculations.
Note that the dynamical equations derived in Sec.~\ref{general} are
completely {\em general}, and could in principle be used
to explore other candidate functional forms for $\Veff(x)$.

In any case, well-documented stand-alone fortran codes for all
three algorithms considered here are available from the author
on demand---which, in the case of (3) at least, will also incorporate
future developments. Note that the weakest algorithmic aspect at present
is the first-order forward time integration of the coupling term
in the dynamical equations---leading to slow $\Dlt$
convergence beyond a certain point, and much smaller typical
$\Dlt$ values than in \Ref{poirier06bohmIII} ($10$ a.u.), which used standard
fourth-order Runge-Kutta integration. The first order of business
is therefore to incorporate a similar time-integrator into
algorithm (3)---or perhaps something even more advanced, such as
the time-adaptive Cash-Carp method,\cite{press} or multistep
Adams-Bashforth method.\cite{phillips}  Preliminary investigations
indicate that this will lead to far greater time step sizes even
than those used in \Ref{poirier06bohmIII}---e.g., $\Dlt = 100$ a.u. or more.
If not the case already, such a modification would result in the
most accurate and efficient quantum dynamics codes available
for 1 DOF systems.

Of course, the primary challenge for future development is
multidimensional systems. As discussed in Sec.~\ref{intro},
the theoretical developments are already in place,
both for stationary state and wavepacket scattering dynamics.
These will be presented in detail in future publications, but
a brief outline may be found in the Summary and Conclusions
of \Ref{poirier06bohmIII}. For the multidimensional stationary state
method, numerical instability of the sort described here
has proven to be an issue. Accordingly, new multidimensional
algorithms will be developed that incorporate the modifications
of the present work. In addition to (hopefully) providing
numerical stability, the reduced number of grid points der
DOF that characterizes the present approach should lead to
far greater reductions in CPU effort for applications with
more DOFs. For wavepacket dynamics, it turns out
that more conventional Bohmian trajectories are appropriate
(defined via $m v_\pm = s_\pm'$). In anticipation of this,
we have analyzed $s_\pm'$ curves for the stationary solutions
obtained here, and found: (1) smooth, but not always monotonic,
behavior; (2) asymptotic agreement with trajectories used here
for all but the $\Pm(x\ra\infty)$ asympotes; (3) exactly
{\em opposite} trajectories in the $\Pm(x\ra\infty)$ asymptote.
The last observation, though curious, may not be significant, as
$r_-(x)$ also vanishes in the same limit.

\begin{acknowledgments}

This work was supported by an award from The Welch Foundation
(D-1523). Corey Trahan is acknowledged for performing some of
the preliminary numerical investigations.
Jason McAfee is also acknowledged for his aid in converting this manuscript to an electronic format suitable for the arXiv preprint server.

\end{acknowledgments}

%
%

%
%
%





%
%

\end{document}